\documentclass{aa}

\usepackage{graphicx}
\usepackage{natbib}
\usepackage{multirow}

\usepackage{txfonts}
\begin{document}

   \title{Contribution of the Galactic center to the local cosmic-ray flux}

   \author{Etienne Jaupart\inst{1} \inst{2} \and Etienne Parizot\inst{3} \and Denis Allard\inst{3}
          }

   \institute{Ecole Normale Sup\'erieure de Lyon, CRAL, UMR CNRS 5574, Universit\'e de Lyon, 69364 Lyon Cedex 07, France\\ email: etienne.jaupart@ens-lyon.fr
        \and D\'epartement de Physique, Ecole Normale Sup\'erieure de Lyon, F-69342 Lyon, France 
         \and APC, AstroParticule et Cosmologie, Universit\'e Paris Diderot, CNRS/IN2P3, CEA/Irfu, Observatoire de Paris, Sorbonne Paris Cit\'e,
75205 Paris Cedex 13, France }

   \date{Received 20 June 2018; accepted 29 July 2018}

 
  \abstract
   {Recent observations of unexpected structures in the Galactic Cosmic Ray (GCR) spectrum and composition, as well as growing evidence for episodes of intense dynamical activity in the inner regions of the Galaxy, call for an evaluation of the high-energy particle acceleration associated with such activity and its potential impact on the global GCR phenomenology.}
   {We investigate whether particles accelerated during high-power episodes around the Galactic center can account for a significant fraction of the observed GCRs, or, conversely, what constraints can be derived regarding their Galactic transport if their contributions are negligible.}
   {
   Particle transport in the Galaxy is described with a two-zone analytical model. We solve for the contribution of a Galactic Center Cosmic-Ray (GCCR) source using Green functions and Bessel expansion, and discuss the required injection power for these GCCRs to influence the global GCR phenomenology at Earth.}
   {We find that, with standard parameters for particle propagation in the galactic disk and halo, the GCCRs can make a significant or even dominant contribution to the total CR flux observed at Earth. Depending on the parameters, such a source can account for both the observed proton flux and boron-to-carbon ratio (in the case of a Kraichnan-like scaling of the diffusion coefficient), or potentially produce spectral and composition features.}
   {Our results show that the contribution of GCCRs cannot be neglected \emph{a priori}, and that they can influence the global GCR phenomenology significantly, thereby calling for a reassessement of the standard inferences from a scenario where GCRs are entirely dominated by a single type of sources distributed throughout the Galactic disk.}

   \keywords{cosmic-rays -- diffusion -- Galaxy: center -- ISM: bubbles -- methods: analytical }

   \maketitle
%
\section{Introduction}

The sources of Galactic cosmic rays (GCRs) remain elusive in spite of decades of intense observational and theoretical efforts. Supernova remnants (\cite{blandford1978particle}, \cite{krymsky1979formation}, \cite{meyer1997galactic}) and superbubbles (\cite{higdon1998cosmic}, \cite{Binns2005}, \cite{bykov1992non}, \cite{parizot2004superbubbles}) have long been acknowledged as promising candidates, based on energy considerations, isotopic composition arguments and a detailed understanding of the characteristics of particle acceleration. Several issues remain outstanding, however, including the $^{22}$Ne signature of GCRs and the maximum energy levels that can be accounted for \citep{lagage1983maximum}. Furthermore, while there is no doubt that these astrophysical environments do accelerate particles, as shown by the high-energy radiation that they generate (\cite{koyama1995evidence}), many questions remain about the magnitude of their actual contribution to the locally observed GCRs. In addition, new observations of unexpected structures in the low-energy GCR spectrum and composition (\cite{pamela2011},  \cite{AMS2015}) raise questions about the respective contributions of different sources in different energy ranges. 

In this context, growing evidence for episodes of intense dynamical activity in the inner regions of the Galaxy (\cite{acero2016development}, \cite{abramowski2016acceleration}) justifies an evaluation of their potential contributions to GCRs and implications for the characteristics of high-energy particle acceleration (\cite{Cheng2012}, \cite{Tibolla2018}). 
Indeed, a total energy release of up to $10^{57}$~ergs has been proposed (\cite{guo2012fermi}), 
which is enough to compete with the average SNR power in the entire Galaxy if the repetition time is of the order of $10^{7}$~years. From a study of how so-called \emph{Fermi bubbles} interact with the Milky Way hot gas halo, \cite{miller2016interaction} have estimated that the average energy injection rate is in a 1--7$\,10^{42}$~erg s$^{-1}$ range, which exceeds the kinetic power due to SN explosions in the interstellar medium. These results have motivated us to investigate 
whether the particles accelerated during these episodes may account for a significant fraction of the GCRs, at Earth and/or elsewhere in the Galaxy, or, conversely, what constraints can be derived about Galactic transport of these particles if their contribution is negligible. In the following, we will refer to these particles as Galactic Center Cosmic Rays (GCCRs).

In this paper, we make a first attempt to address these important questions by studying the contribution of a continuous source of energetic particles at the center of the Galaxy to the local GCRs. Our calculations rely on a simplified propagation model similar to that which is used in generic studies of GCR phenomenology (\cite{ginzburg2013origin}, \cite{Strong1998}, \cite{taillet2003spatial}, \cite{Bringman2007}, \cite{boudaud2015fussy}, \cite{Giesen2015}, \cite{genolini2015}). This model includes energy-dependent diffusion and advection in a Galactic wind, energy losses and particle re-acceleration, and is described in Sect.~\ref{sec:model}. The formalism and resolution scheme are presented in Sect.~\ref{sec:ResolutionScheme} and results are shown in Sect.~\ref{sec:Results}. A summary and discussion of the results are proposed in Sect.~\ref{sec:discussion}.

\section{Model description}
\label{sec:model}

\begin{figure}[!h]
\centering
\includegraphics[width=0.35\textwidth]{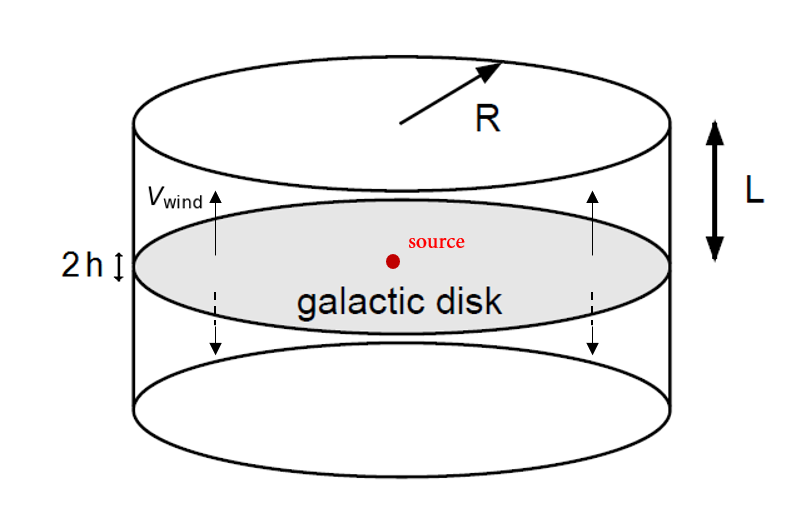}
\caption{Geometry of the diffusive volume.}
\label{fig:geom}
\end{figure}
For the present exploratory study, we use a classical, simplified, two-zone model of the Galaxy, consisting of a cylindrical homogeneous disk with radius $R = 20$~kpc and half-thickness $h = 100$~pc, surrounded by a cylindrical magnetic halo with the same axis and radius, $R$, and with a half-thickness $L \gg h$ (see e.g. \cite{taillet2003spatial}). Cosmic-ray transport is treated through a standard diffusion equation (see e.g. \cite{Strong1998}, \cite{taillet2003spatial}), including a convective term corresponding to a wind in the $z$ direction orthogonal to the Galactic disk, dragging cosmic rays away from the disk with a velocity $V_\mathrm{wind}$. Diffusion is assumed to be isotropic and homogeneous over the entire ``confinement volume''. Diffusion coefficient $D$ depends on the magnetic rigidity of the particles (see e.g. \cite{Bringman2007}) :
\begin{equation}
D = D_{0} \, \beta \, R_{\mathrm{GV}}^{\delta}, 
\end{equation}
where $\beta = v/c$, $R_{\mathrm{GV}}$ is the rigidity in units of GV, $D_{0}$ is a constant expressed in kpc$^{2}$/Myr,and $\delta$ depends on the type of turbulence underlying the diffusion process. Here, we allow for both Kolmogorov and Kraichnan turbulence spectra, corresponding to $\delta = 1/3$ and $\delta = 1/2$, respectively.

It is convenient to use the kinetic energy of the particles, $K$, as the primary variable. 
The particle spectral density at energy $K$, $\Psi(K, \vec{r},t)$ is written as a function of time $t$ and position with respect to the Galactic center $\vec{r}$. Neglecting the spallation of heavier species, it must satisfy the following diffusion equation:
\begin{eqnarray}
\frac{\partial}{\partial t}\Psi - D(K)\Delta \Psi + \frac{\partial}{\partial z} (V_\mathrm{wind} \Psi) \nonumber \\
+ \frac{\partial}{\partial K} [b_{\mathrm{loss}}(K) \Psi - \chi(K) \frac{\partial}{\partial K} \Psi ]  = Q(K,\vec{r},t) - \Gamma_{\mathrm{ISM}}\Psi,
\label{eq:DiffusionEquation}
\end{eqnarray}
where $Q(K,\vec{r},t)$ is a source term to be specified, $\Gamma_{\mathrm{ISM}}$ is the rate of ``catastrophic losses'' of the particles (decay or destruction due to interactions with ambient matter in the interstellar medium), $b_{\mathrm{loss}}(K)$ is the rate of energy loss, and $\chi(K)$ is an effective diffusion coefficient in energy space, associated with a re-acceleration process accompanying the diffusion of particles in space. In its simplest form, the latter is traditionally modeled as a Fermi second-order acceleration process related 
to the ambient magnetic turbulence, which is assumed to be isotropic, and can be expressed in terms of a single parameter, namely the Alfv\'en speed, $v_{\mathrm{A}}$. In this model, the diffusion in energy space is thus inseparable from that in geometric space, and the coefficient $\chi(K)$ can in practice be related to $D(K)$ by (see \cite{maurin2001cosmic}):
\begin{equation}
\chi(K) = \frac{2}{9}v_{\mathrm{A}}^{2}\frac{\beta^{4}E_{\mathrm{tot}}^{2}}{D(K)},
\label{eq:chiE}
\end{equation}
where $E_{\mathrm{tot}} = K + mc^{2}$.

Energy loss occurs due to ionization and adiabatic processes, such that $b^{\mathrm{loss}} = b^{\mathrm{ion}} +  b^{\mathrm{adiab}}$. We further consider that the interstellar matter only fills the ``infinitely thin'' disk (with respect to other dimensions) and use cylindrical coordinates centered on the galactic axis. Assuming cylindrical symmetry, the energy loss term in Eq.~(\ref{eq:DiffusionEquation}) may be written as
\begin{equation}
b_{\mathrm{loss}}(K)\Psi(K,\vec{r},t) = b_{\mathrm{loss}}(K)\Psi(K,r,z=0,t)\times 2h\delta(z),
\label{eq:lossTerm}
\end{equation}
where $r$ is the galactocentric radius: $\vec{r} = r\,\vec{u}_{r} + z\,\vec{u}_{z}$.
If the ``discofugal'' wind reaches its nominal velocity $V_{\mathrm{wind}}$ at the top and bottom of the disk, the overall effect of adiabatic losses can be written as:
\begin{equation}
b_{\mathrm{adiab}} = - \frac{V_{\mathrm{wind}}}{3h}\frac{p^{2}}{E_{\mathrm{tot}}}.
\label{eq:bAdiab}
\end{equation}

It has been also argued by  \cite{ptuskin1997} that re-acceleration may be mostly efficient within the disk, because it would not work in the adopted wind model. We do not believe that this should be necessarily the case, since some degree of re-acceleration should also accompany diffusion, even in the case of a non isotropic turbulence. However, this assumption was also adopted by \cite{maurin2001cosmic}, \cite{taillet2003spatial}, \cite{Donato2004} and \cite{Giesen2015}  in their study of GCR propagation, and we shall use it here for the sake of simplicity and to allow a direct comparison of our results. We thus write:
\begin{equation}
\chi(K)\frac{\partial \Psi}{\partial K}(K,\vec{r},t) =\chi(K)\frac{\partial \Psi}{\partial K}(K,r,z=0,t)\times 2h\delta(z),
\label{eq:ReaccTerm}
\end{equation}

Likewise, the source term can be written $Q(K,\vec{r},t) = Q(K,r,z=0,t) \times 2h\,\delta(z)$ if the sources are distributed in the Galactic disk, and
\begin{equation}
Q(K,\vec{r},t) = \frac{\mathrm{d}N}{\mathrm{d}K}(K)\,f(t)\delta(\vec{r}),
\label{eq:sourceTerm}
\end{equation}
for a central source, where $f(t)$ is a function of time allowing for time changes of the particle injection rate. For $\mathrm{d}N/\mathrm{d}K$, we take a power law in momentum (in some relevant energy range), with logarithmic index $\alpha$: $\mathrm{d}N/\mathrm{d}p = N_{0}(p/p_{0})^{-\alpha}$. The normalization of this injection term is directly related to the total injection power from that source:
\begin{equation}
\mathcal{P}_{\mathrm{inj}}(t) = f(t)\times \int K\,\frac{\mathrm{d}N}{\mathrm{d}K}(K)\,\mathrm{d}K.
\label{eq:sourcePower}
\end{equation}

Boundary conditions follow from the standard assumption that the magnetic halo has finite dimensions, such that diffusion only confines cosmic rays within a limited volume. Outside this ``volume'', particles escape freely with a velocity close to $c$, resulting in a practically vanishing density. 
We thus impose:
\begin{equation}
\Psi(K,r=R,z,t)=\Psi(K,r,z=\pm L,t)=0.
\end{equation}


\section{Resolution scheme}
\label{sec:ResolutionScheme}
\subsection{Discrete Fourier-Bessel expansion}

With these boundary conditions, Eq. (\ref{eq:DiffusionEquation}) can be solved in cylindrical coordinates by expanding the cosmic-ray spectral density as a series of Bessel functions (see e.g. \cite{eg1953morse}):
\begin{equation}
\Psi(K,r,z,t) = \sum_{i=1}^{+ \infty} P_i(K,z,t) \times J_0(u_i \frac{r}{R}) \equiv \sum_{i=1}^{+ \infty} \Psi_{i}(K,r,z,t),
\label{eq:PsiSum}
\end{equation}
where $\{u_i\}_{i\geq 1}$ are the zeros, ranked by increasing order, of the zeroth-order Bessel function of the first kind, $J_{0}(x)$. $J_{0}(u_{i}r/R)$ is an eigenfunction of the Laplacian operator appearing in the diffusion equation, with eigenvalue $-u_{i}^{2}/R^{2}$. Thus:
\begin{equation}
   \Delta \Psi_{i}(K,r,z,t) = \left[\frac{\partial^{2}}{\partial z^{2}} P_{i}(K,z,t) - \frac{u_{i}^{2}}{R^{2}} P_{i}(K,z,t) \right] \times J_{0}(u_{i}r/R).
\end{equation}
Applying the Fourier-Bessel transform, $\int_{0}^{R}rf(r)J_{0}(u_{i}r/R)\mathrm{d}r = f_{i}\times J_{1}(u_{i}^{2})R^{2}/2$, to Eq.~((\ref{eq:DiffusionEquation})), one derives equations for each Bessel ``coefficient'' of order $i$, $P_i(K,z,t)$:

\begin{eqnarray}
\frac{\partial}{\partial t}P_{i} - D \frac{\partial^{2}P_{i}}{\partial z^{2}} + D \frac{u_{i}^{2}}{R^{2}}P_{i}+  \frac{\partial}{\partial z} (V_\mathrm{wind} P_{i}) \nonumber \\
+ \frac{\partial}{\partial K}[b_{\mathrm{loss}} P_{i} - \chi \frac{\partial}{\partial K} P_{i}]  = Q_{i}(K,z,t) - \Gamma_{\mathrm{ISM}} P_{i},
\label{eq:Bessel}
\end{eqnarray}
where $Q_{i}(K,z,t)$ is the $i^{\mathrm{th}}$ order Bessel coefficient of the source term, which writes, for a source term as in Eq.~(\ref{eq:sourceTerm}):
\begin{equation}
Q_i(K,z,t)=f(t) \, \frac{\mathrm{d}N/\mathrm{d}K}{\pi R^2 J_1(u_i)^2} \,\delta(z).
\label{eq:sourceTermBessel}
\end{equation}

By symmetry, $\Psi(K,r,z,t)$ is expected to be an even function of $z$, and so will be $P_{i}(K,z,t)$, $\forall i$.

\subsection{Steady-state solution}

In this paper, we concentrate on steady-state solutions, such that $\partial P_{i}/\partial t = 0$ in Eq.~(\ref{eq:Bessel}). The resolution follows standard procedures, as in \cite{taillet2003spatial} for example. We first solve Eq.~(\ref{eq:Bessel}) in the halo, $i.e.$ outside the disk: $z \neq 0$, where both the right-hand side of Eq.~(\ref{eq:Bessel}) 
and the energy loss/re-acceleration term vanish. Restricting ourselves to $z > 0$, where $V_{\mathrm{wind}}$ is constant, we rewrite the equation as follows (after dividing by $-D(K)$):
\begin{equation}
\frac{\partial^{2}P_{i}}{\partial z^{2}} - \frac{V_\mathrm{wind}}{D(K)} \frac{\partial P_{i}}{\partial z} - \frac{u_{i}^{2}}{R^{2}}P_{i} = 0,
\label{eq:BesselPrime}
\end{equation}
Integrating with the boundary conditions $P_{i}(z = L) = 0$ yields:
\begin{equation}
P_i(K,z) = P_i(K,0) \times \exp \left(\frac{V_\mathrm{wind}}{2D(K)} z \right) \times \frac{\sinh(S_i(L-z))}{\sinh(S_iL)},
\label{eq:PiStationary}
\end{equation}
where $S_i=[(V_\mathrm{wind}/2D(K))^2 + (u_i/R)^2]^{1/2}$.

To obtain the solution in the disk plane, $P_i(K,0)$, one integrates Eq.~(\ref{eq:Bessel}) through the disk between $[-\epsilon,\epsilon]$ and take the limit as $\epsilon \to 0$. This gives:
\begin{eqnarray}
2 V_\mathrm{wind} P_i(K,0) - 2 D(K) \frac{\partial}{\partial z} P_i(0^+)  \nonumber \\
+ 2h \frac{\partial}{\partial K} [ b_{\mathrm{loss}}P_i(K,0) - \chi(K) \frac{\partial}{\partial K} P_i(K,0) ] \nonumber \\
= f(t)\frac{\mathrm{d}N/\mathrm{d}K}{\pi R^2 J_1(u_i)^2} -2h\Gamma_{\mathrm{ISM}}(K)P_{i}(K,0),
\label{eq:Pi0}
\end{eqnarray}
where $f(t) = 1 \,\mathrm{s}^{-1}$ in steady-state conditions. For non-radioactive primary particles, $\Gamma_{\mathrm{ISM}} = n_{\mathrm{ISM}} \, \sigma_{\mathrm{D}}(K) \,v$, where $v$ is the particle velocity and $\sigma_{\mathrm{D}}(K)$ the total destruction cross section for that particle due to interactions in the ISM of homogeneous density $n_{\mathrm{ISM}}$.

Equation~(\ref{eq:Pi0}) is then solved by discretizing in energy space, as detailed in the Appendix, providing $P_i(K,0)$ and thus $P_i(K,z)$ for all $i$, from which the particle spectral density is finally obtained at all positions $(r,z)$ by summing over a sufficiently large number of terms (see Eq.~\ref{eq:PsiSum}). 
In practice, the truncated series $S_{N} = \sum_{i=1}^{N} \Psi_{i}(K,r,z)$, is slowly oscillating when $N$ increases, and we found that accelerated convergence is ensured by computing the average of a large enough number of terms, beyond a certain order. Here, we used for most calculations $\Psi(K,r,z) \simeq (\sum_{n = 1}^{300}S_{2000+n})/300$.

\subsection{Secondary particles}
\label{subsec::secondary}
The above approach can easily be extended to compute the distribution of secondary particles, 
produced in flight by the interactions of the primary cosmic rays with the ISM, through the standard spallation process. The formalism remains the same with a source term appropriate for spallation. Noting $\sigma_{\mathrm{I}+\mathrm{T}\rightarrow\mathrm{II}}(K^{\prime},K)$ the differential cross section for the production of secondary nuclei II at energy $K$, by the interaction of a primary cosmic ray of type I of energy $K^{\prime}$ with a target nucleus of type T with density $n_{\mathrm{T}}^{(\mathrm{ISM})}$, and summing over all spallation channels, 
the source term for nuclei S is given by:
\begin{equation}
Q^{(\mathrm{II})}(K,r,z) = \sum_{\mathrm{I},\mathrm{T}}\,n_{\mathrm{T}}^{(\mathrm{ISM})}\times \int \Psi^{(\mathrm{I})}(K^{\prime},r,t)\sigma_{\mathrm{I}+\mathrm{T}\rightarrow\mathrm{II}}(K^{\prime},K)v(K^{\prime})\mathrm{d}K^{\prime}.
\label{eq:spallation}
\end{equation}

For our present purposes, we approximate this source term by assuming that spallation products in the cosmic rays keep the same energy per nucleon as its energetic progenitor, and consider only protons in the ISM, with an average density $n_{\mathrm{ISM}} \simeq 1.3\,\mathrm{cm}^{-3}$, with cross sections taken from \cite{silberberg1973partial}.

This leads to the Fourier-Bessel coefficient of order $i$ for the source term of secondary nuclei of type II:
\begin{equation}
Q_{i}^{(\mathrm{II})}(K,r,z,t) = 2h\delta(z) \times \sum_{I} n_{\mathrm{ISM}}\sigma_{\mathrm{I}\to \mathrm{II}}(K \frac{A_{\mathrm{I}}}{A_{\mathrm{II}}}) v(K\frac{A_{\mathrm{I}}}{A_{\mathrm{II}}}) P_{i}^{(\mathrm{I})}(K\frac{A_{\mathrm{I}}}{A_{\mathrm{II}}}),
\label{eq:secondarySources}
\end{equation}
where $A_{\mathrm{I}}$ and $A_{\mathrm{II}}$ are the atomic mass numbers of the primary and secondary particles, respectively.

\begin{table*}[!ht]
\caption{Parameters for the models used in this study as well as the injected power obtained by matching the calculated flux with the observed one. The number appearing in the name of the model corresponds to the value of the logarithmic exponent $\alpha$ in the source term.}             
\label{table:power}      
\centering                          
\begin{tabular}{|c |c |c |c |c |c |c || c|}        
\hline                 
Model & $L$ [kpc] & $D_0$ [kpc$^2$ Myr$^{-1}$] & $\delta$ & $V_\mathrm{wind}$ [km/s] & $v_\mathrm{A}$ [km/s] & $\alpha$ & $\mathcal{P}_{\mathrm{inj}}$ [erg/s]  \\    
\hline                        
\hline
   MIN2.4 & 1.0 & 0.0016 & 0.85 & 13.5 & 22.4 & 2.4 & 2.9 10$^{45}$ \\
   \hline
   MED2.2 & \multirow{3}{*}{4.0} & \multirow{3}{*}{0.012} & \multirow{3}{*}{0.7} & \multirow{3}{*}{12.0} & \multirow{3}{*}{52.9} & 2.2 & 3.2 10$^{41}$ \\
   MED2.3 & & & & & & 2.3 & 4.3 10$^{41}$\\
   MED2.4 & & & & & & 2.4 & 6.5 10$^{41}$ \\
   \hline
   MAX2.2 & \multirow{3}{*}{15.0} & \multirow{3}{*}{0.0765} & \multirow{3}{*}{0.46} & \multirow{3}{*}{5.0} & \multirow{3}{*}{117.6} & 2.2 & 6.9 10$^{40}$ \\
   MAX2.3 & & & & & & 2.3 & 9.2 10$^{40}$ \\ 
   MAX2.4 & & & & & & 2.4 &  1.4 10$^{41}$ \\
   \hline
   *Kr2.4 & 10 & 0.09 & 0.5 & 10.0 & 28.0 & 2.4 & 2.8 10$^{41}$ \\
   \hline
   *Kol2.55 & 10 & 0.23 & 1/3 & 12.0 & 30.0 & 2.55 & 6 10$^{41}$\\
   
\hline                                   
\end{tabular}
\tablefoot{*Models with a star are described in section \ref{subsec:InfluenceParam} and reproduce the observed primary fluxes}
\end{table*}

\subsection{Solar modulation}

To fully describe CR transport to Earth, one has to include the influence of the Sun for the very last part of their flight. Close to Earth, CRs penetrate the Sun's sphere of influence and are subjected to a phenomenon called Solar Modulation ('Smod'). The solar wind and associated magnetic field significantly reduce the kinetic energy of low energy CR ($T \lesssim  10$ GeV/n) 
and prevent these from reaching our planet. This effect can be effectively described by a Fisk potential $\Phi_F$ in the 'force field approximation'. The flux in the local interstellar environment (LIS) is modulated to obtain the flux on Earth $\Phi_{\oplus} (K)$ 
as follows (\cite{gleeson1968solar}, \cite{boudaud2015fussy})
\begin{equation}
\Phi_{\oplus} (K)= \Phi_{LIS} (K + |e| \Phi_F Z/A) \times \frac{K(K+2m)}{(K+m+|e|\Phi_F Z/A)^2 - m^2}
\end{equation}

For Pamela data, conservative estimates of $\Phi_F$ are $0.1 \, \text{GV} < \Phi_F < 1.1 \, \text{GV}$ and for AMS $0 \, \text{GV} < \Phi_F < 2 \, \text{GV}$. 

\section{Results} 
\label{sec:Results}
\subsection{Typical expectations} 
\label{subsec:Expectation}

To evaluate contributions from episodes of intense activity at the Galactic center, we first evaluate the power that needs to be injected in GCCRs in these episodes to obtain local CR fluxes comparable to those observed at Earth. 

We first use benchmark values for the propagation parameters, taken from models that reproduce the observed secondary-to-primary ratios with a distribution of primary sources corresponding to supernovae remnants (\cite{Case1998}, \cite{maurin2001cosmic}). These models are called MIN, MED and MAX (for minimum, medium and maximum, referring primarily to the thickness of the halo) 
and are referenced in table \ref{table:power}. To obtain the required power injected as CRs at the Galactic center (see Eq.~\ref{eq:sourcePower}), we specify the injection rate $\mathrm{d}N/\mathrm{d}K$ by specifying values for the logarithmic index $\alpha$ and the normalization coefficient $N_0$ in the momentum power law. We choose values between $2.2$ and $2.4$ \citep{achterberg2001particle} for the former, and adjust the value of the latter numerically so that the calculated flux matches the data at high energy (here $K=600$ GeV).Thus, coefficient $N_0$ is just a scaling factor which does not affect trends in the results. The data are taken 
from the Cosmic Ray Database (\cite{maurin2014database} and references therein).

We find that for the MIN model the required injected power is much larger than that available, due to the small halo size, $L$ (see section \ref{subsec:InfluenceParam}). However, recent studies based on synchrotron radio emission (\cite{di2013cosmic}, \cite{fornengo2014isotropic}) but also positrons (\cite{boudaud2015new}, \cite{lavalle2014direct}) and anti-protons (\cite{Giesen2015}), do not support the thin halo of MIN models. For the other two benchmark models (MED and MAX), a fraction of the global power budget from high energy events 
at the Galactic center (\cite{guo2012fermi}, \cite{miller2016interaction}) is sufficient to match the observed flux.

\begin{figure}[!ht]
\includegraphics[width=\hsize]{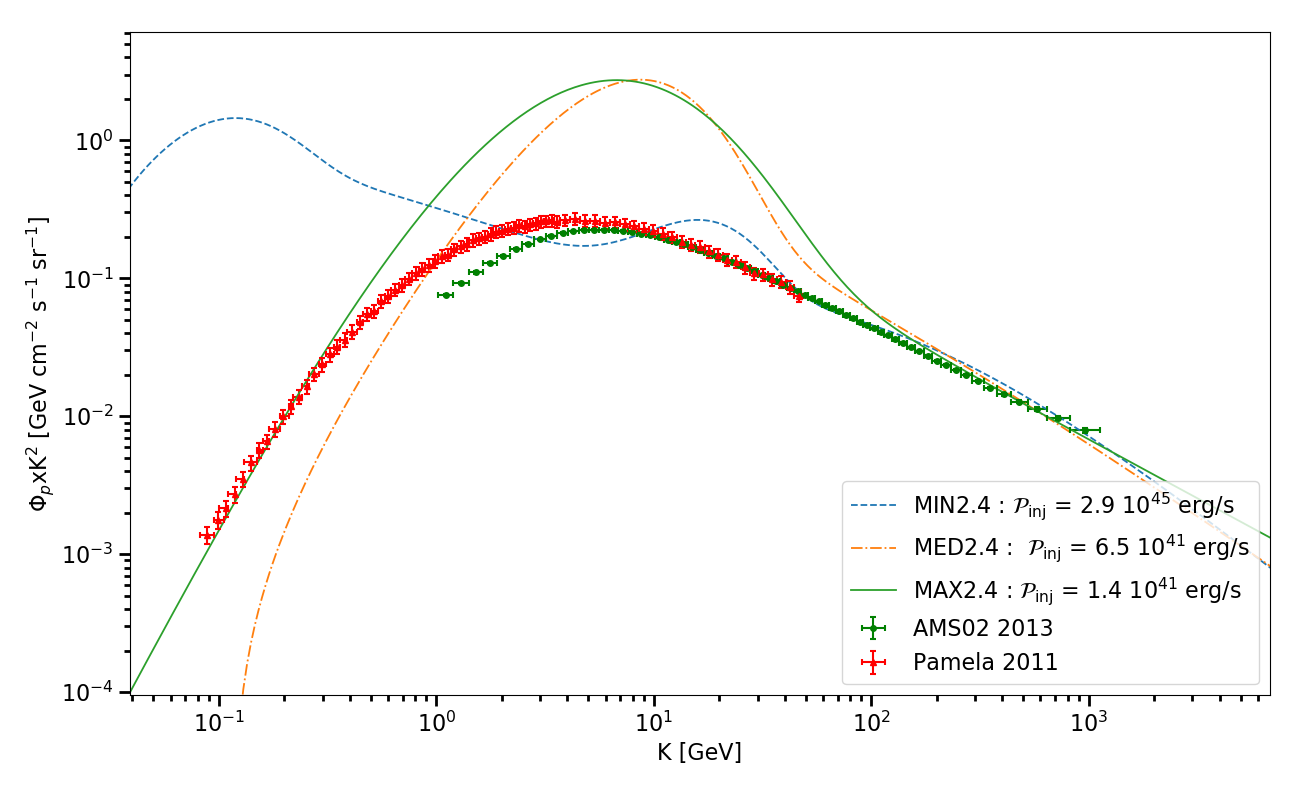}
\caption{H protons flux $\Phi_p$ at Earth, rescaled by $K^2$, for the three standard models (at the obtained injection powers) 
against kinetic energy $K$. Data from Pamela and AMS02 are also displayed.}
\label{fig:compModel}
\end{figure}

Table \ref{table:power} gives a list of the parameters of the different models and the corresponding injection power, as well as the parameters and injection power for the models of section \ref{subsec:InfluenceParam}. Figure \ref{fig:compModel} shows the cosmic-ray spectral densities at Earth obtained from simulations for the three benchmark models, assuming the indicated injection power, compared to the observed flux.

From Fig. \ref{fig:compModel}, it appears that with only 10\% of the matching injection power (downward shift of the curves by one decade), Galactic center bursts can still be expected to generate features in the observed flux. With the parameters of the MED or MAX models, a CR injection power of $\sim$1--6$\, 10^{40}$ erg/s, i.e. merely a few percent at most of the total energy injection rate of 1--7$\, 10^{42}$ erg/s of events leading to Fermi Bubbles (\cite{guo2012fermi}, \cite{miller2016interaction}), is sufficient to make a significant contribution to the locally observed GCR fluxes, at least in some energy range.Note that these total injection powers in GCRs are close to those that are derived from models involving a classical source distribution  (\cite{strong2010global}).

\begin{figure*}[ht!]
\centering
\includegraphics[width=0.49\textwidth]{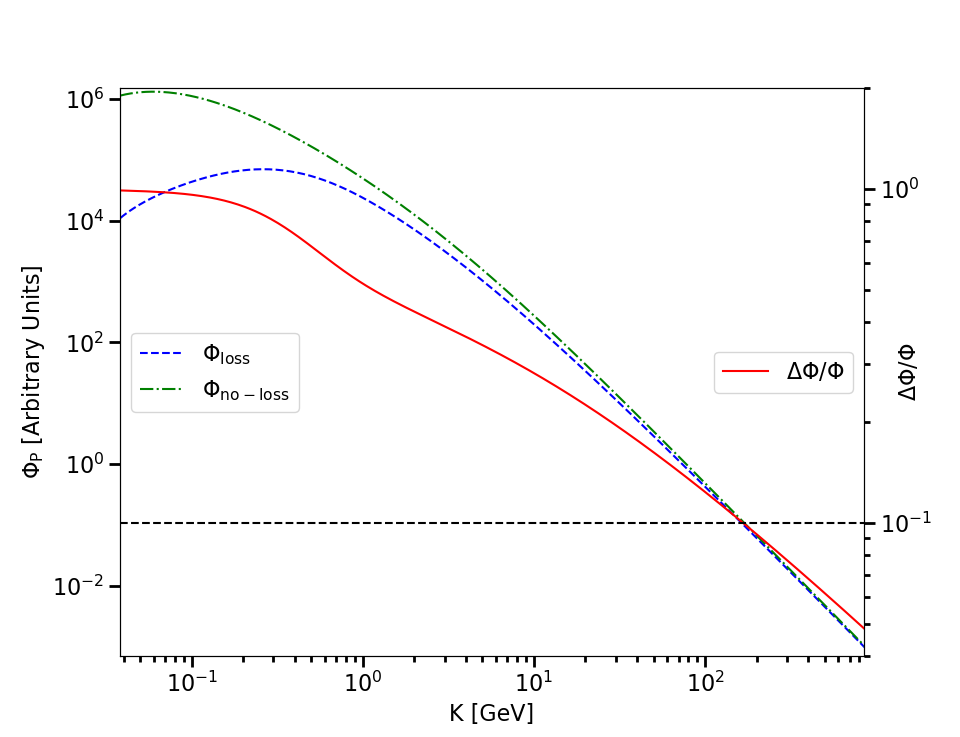}
\includegraphics[width=0.49\textwidth]{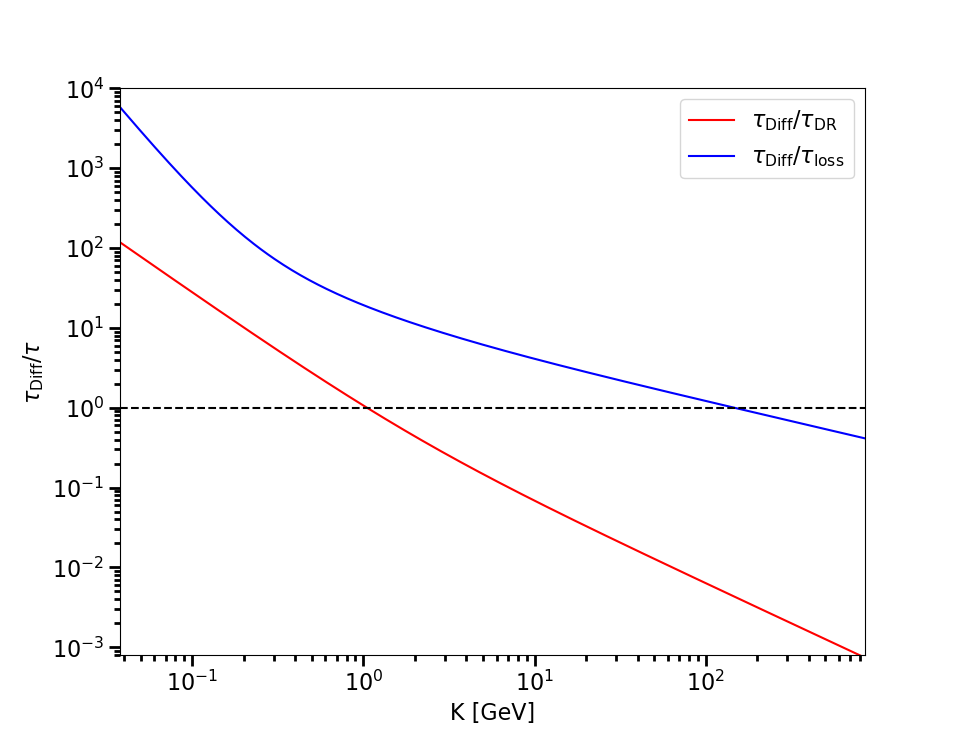}
\caption{(Left) H protons flux $\Phi_p$ at Earth, for the Kr2.4 model with and without energy losses and re-acceleration against kinetic energy ($K$). Solar modulation was turned off in both cases, such that $\Phi_F=0$. (Right) Timescales ratios against kinetic energy for the Kr2.4 model.}
\label{fig:losses}
\end{figure*}

From this simple estimate, one may conclude that a global model of the local GCRs should not be limited to standard sources distributed throughout the Galaxy and should also include the contribution of GCCRs. For there are only two alternatives: the contribution of these cosmic rays is negligible or significant. In the former case, one must understand why the high energy sources that are present in the Galaxy center do not contribute much to the local GCRs. This could be due to a very inefficient conversion of the available power into energetic particles or to an unexpectedly thin halo, but this would likely conflict with previous arguments (\cite{di2013cosmic}, \cite{fornengo2014isotropic}, \cite{boudaud2015new}, \cite{lavalle2014direct}). If the contribution of GCCRs is significant, the ``standard parameters" derived for the propagation of GCRs should be modified accordingly. As shown below, GCCRs might even account for the majority of the observed GCRs for some particular parameter choices, which emphasizes further that common results on GCR propagation rely on specific assumptions regarding the acceleration and transport of energetic particles from high-power events in the central part of the Galaxy.

\subsection{Influence of parameters}
\label{subsec:InfluenceParam}

\subsubsection{Energy losses and re-acceleration}
\label{subsubsec:loss}

Energy losses and re-acceleration play a role in the low energy range of the GCRs spectrum. The typical timescales for energy losses, $\tau_{\mathrm{loss}} = - K/b_{\mathrm{loss}}(K)$, and diffusive re-acceleration, $\tau_{\mathrm{DR}} = K^2/\chi(K)$, can be compared with the typical confinement timescale either derived from a one dimensional slab model or from the Leaky Box model (e.g. \cite{boudaud2015fussy}). In these models involving a homogeneous distribution of sources, both energy losses and re-acceleration become negligible at kinetic energies above of a few GeV. 

In the case of a central source, however, the relevant timescale is the typical time needed to reach the Earth from the central source by diffusion $\tau_\mathrm{Diff}$, which is of the order of $r_{\odot}^2/2D(K)$, where $r_{\odot} \approx 8.5$ kpc the distance of the Solar-system to the Galactic center. These timescales are represented in Fig.~\ref{fig:losses} (right). We find that energy losses are significant up to $\sim 10^2$ GeV. This can also be verified with simulations by turning on and off the energy losses and diffusive re-acceleration. Figure \ref{fig:losses} (left) shows the resulting fluxes for the Kr2.4 model (see table \ref{table:power}), with arbitrary normalization. Also shown is the relative difference $\Delta \Phi / \Phi = |\Phi_{\mathrm{loss}} - \Phi_{\mathrm{no-loss}}|/\Phi_{\mathrm{loss}}$, which can be seen to drop below 10\% above $\sim 150$~GeV in this configuration. At higher energies, parameters $V_\mathrm{wind}$ and $v_\mathrm{A}$ have essentially no incidence on the cosmic ray flux, whose level is then controlled by the other parameters.

\subsubsection{Halo size}
\label{subsubsec:halosize}

\begin{figure}[ht]
\centering
\includegraphics[width=\hsize]{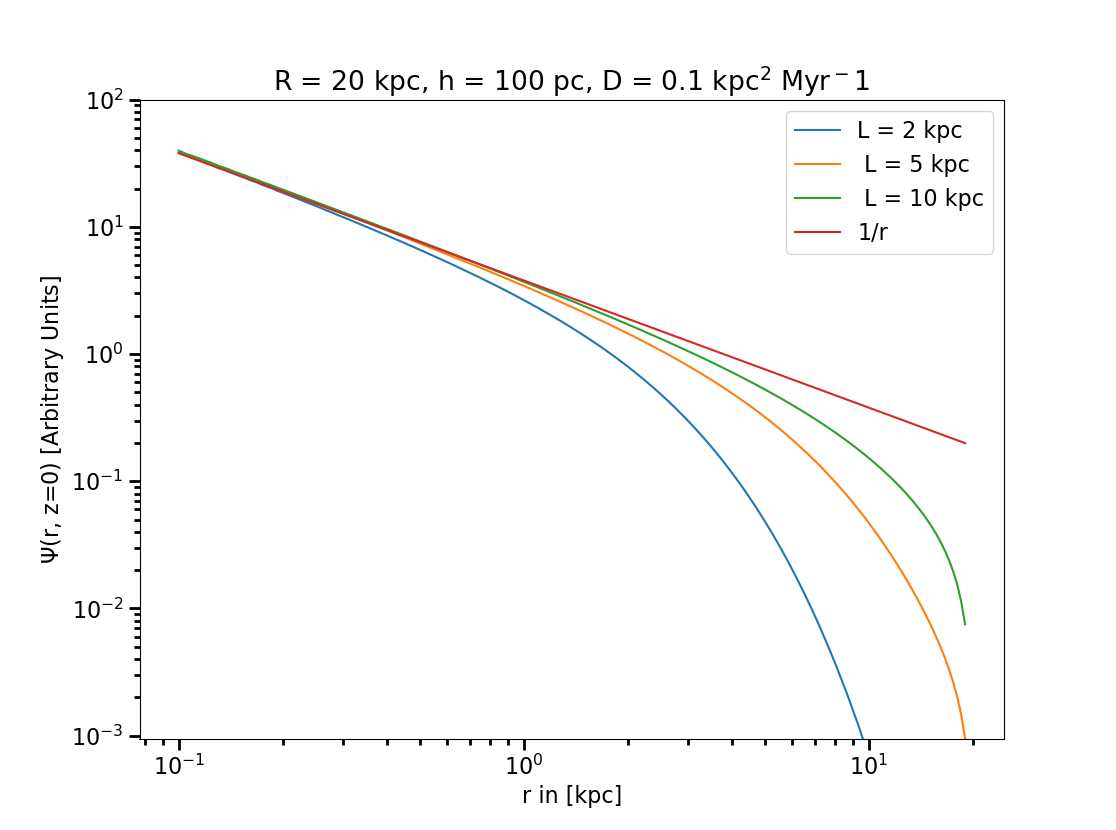}
\caption{Density profiles for a various set of halo size $L$ for a purely diffusive equation. The profile for an infinite box (in both direction) is also displayed and is labeled $1/r$ as $\Psi = N_0/4\pi D r$ in this case.}
\label{fig:halo}
\end{figure}

As mentioned in section \ref{subsec:Expectation}, the halo size has a large influence on the power that needs to be injected to obtain CR fluxes comparable with the ones observed. More particles indeed escape the halo before arriving in the Earth's vicinity as the confinement box gets smaller. This can be seen within a purely diffusive model, which is a good approximation of our complete model at high energy (see Sect.~\ref{subsubsec:loss}). Figure \ref{fig:halo} shows the corresponding density profiles in the Galactic disk as a function of the galactocentric radius $r$, for various sizes of the halo, $L$, and for a fixed diffusion coefficient $D$ (and thus energy $K$). As can be seen, at a distance $r_\odot \approx 8.5$ kpc, the size of the halo becomes critical around $L \leq 2$ kpc: for lower values of $L$, the local spectral density $\Psi$ drops significantly below from that obtained for $L=5$~kpc, $L=10$~kpc or for an infinite box case. This explains why a very large injection power is needed to compete with locally observed GCR flux in benchmark model MIN with a thin halo.

\subsubsection{Degeneracy in the choice of parameters}

\begin{figure}[ht!]
\includegraphics[width=\hsize]{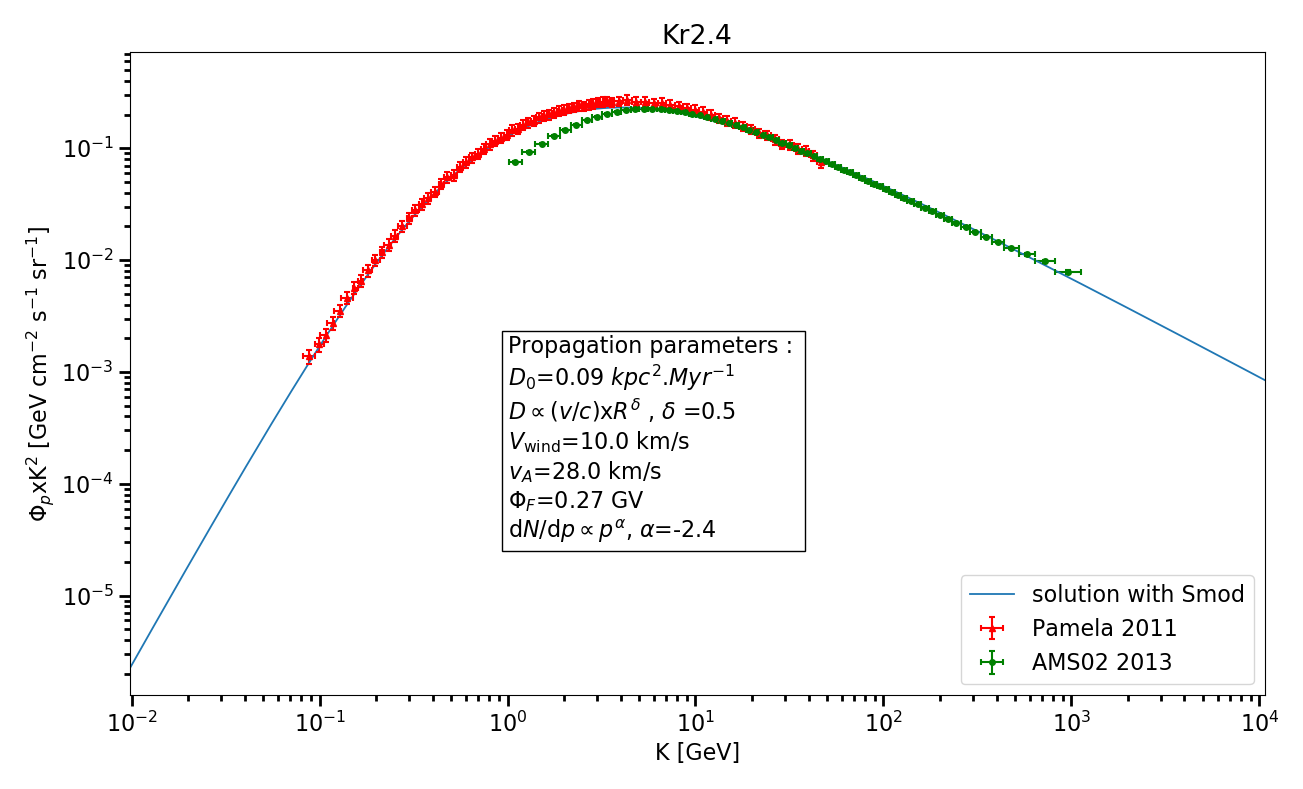}
\caption{H protons flux $\Phi_p$ at Earth, rescaled by $K^2$, for the Kr2.4 model (see table \ref{table:power}) against kinetic energy $K$ }
\label{fig:KR24}
\end{figure}

\begin{figure}[ht!]
\includegraphics[width=\hsize]{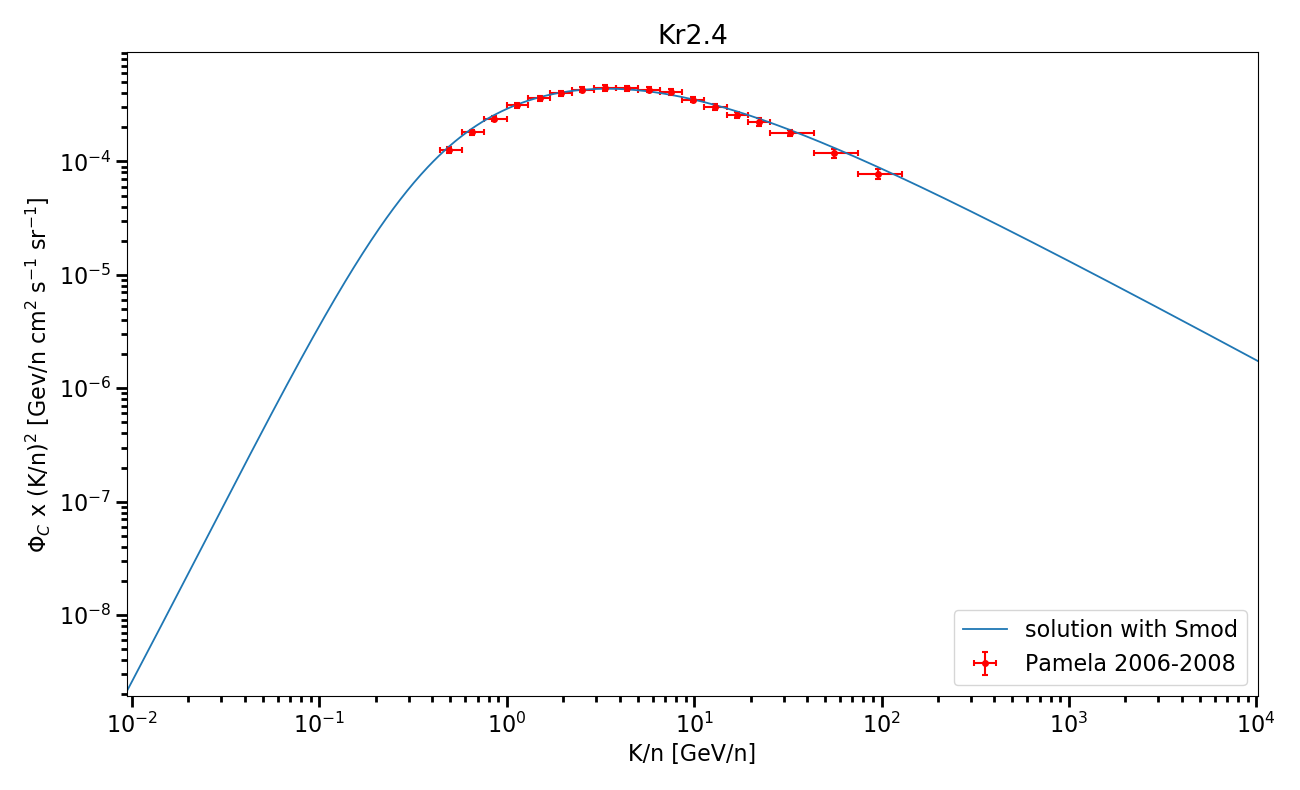}
\caption{Carbon flux $\Phi_C$ at Earth, rescaled by $(K/n)^2$, for the Kr2.4 model (see table \ref{table:power}) against kinetic energy per nucleon $K/n$.}
\label{fig:KR24C}
\end{figure}
By solving Eq.~(\ref{eq:Pi0}) without energy losses and re-acceleration (which only influence the low-energy part of the spectrum), one may write $\Psi(K,r,z=0) = N_0 K^{-\alpha}/D(K) \times g(r,R,h,L)$, where $g$ is a function of spatial variables only. Consequently, the power injected as GCCRs, $\mathcal{P}_\mathrm{inj}$, required to produce CR fluxes comparable to the observed ones at Earth is (in our model) a function of the halo size $L$, the diffusive coefficient $D$ and the spectral index $\alpha$. However, there is some degeneracy in the choice of parameters, similar to that expressed by the three benchmark models. More precisely, a negligible influence of the activity at the Galactic center on the observed GCRs can be obtained by invoking either an unusually thin halo or a particularly large diffusion coefficient for example (since $\Psi \propto 1/D(K)$ at high energy). Conversely, if a source at the Galactic center turns out to have a significant impact on the observed fluxes, some degeneracy must be accepted at this stage between $L$ and $D$.

With this caveat in mind, we investigated whether it could even be possible to account for the entire GCR spectrum with a single central source. Figures \ref{fig:KR24} and \ref{fig:KR24C} show flux values that are obtained for the Kr2.4 model, i.e. a model with a source spectrum in $p^{-2.4}$ and a Kraichnan-like turbulence scaling for the energy diffusion coefficient (see above). With the parameters given in Table \ref{table:power}, we find that an injection power $\mathcal{P}_\mathrm{inj} = 2.8 \,10^{41}$~erg/s allows an excellent fit  to the observed flux. 
Such a power appears accessible in principle, given the average power range of 1--7 $10^{42}$ erg/s available from high energy events at the Galactic center.

\begin{figure}[ht!]
\includegraphics[width=\hsize]{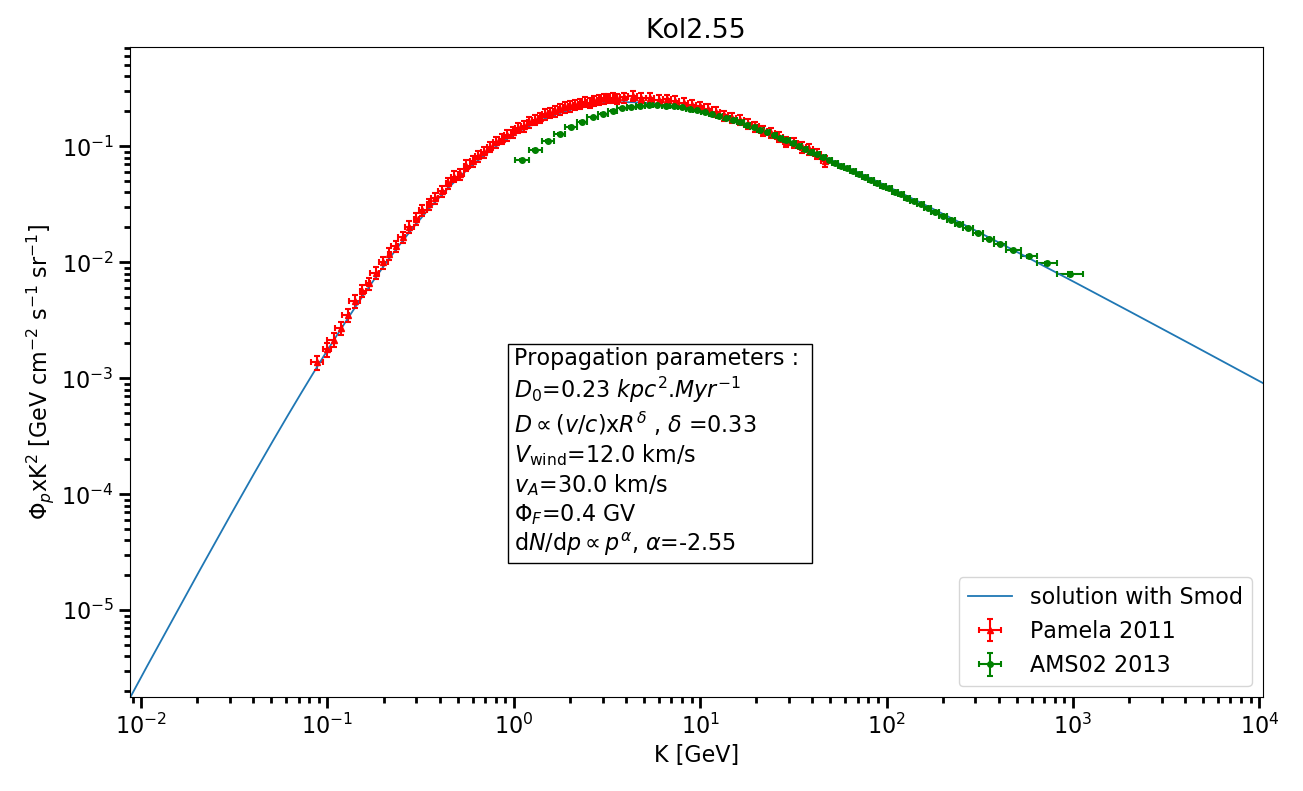}
\caption{H protons flux $\Phi_p$ at Earth, rescaled by $K^2$, for a Kolmogorov model, against kinetic energy $K$. Propagation parameters are displayed and we obtain an injection power $\mathcal{P}_\mathrm{inj} = 6 \, 10^{41}$ erg/s.}
\label{fig:Kl255}
\end{figure}

Other sets of parameters can also reproduce the data for reasonable values of the injection power. An example is shown on Fig. \ref{fig:Kl255}, with a model using a Kolmogorov-like diffusion coefficient ($\delta = 1/3$) and a halo size $L=10$ kpc. All the other parameters are also displayed.

\begin{figure}[!h]
\includegraphics[width=\hsize]{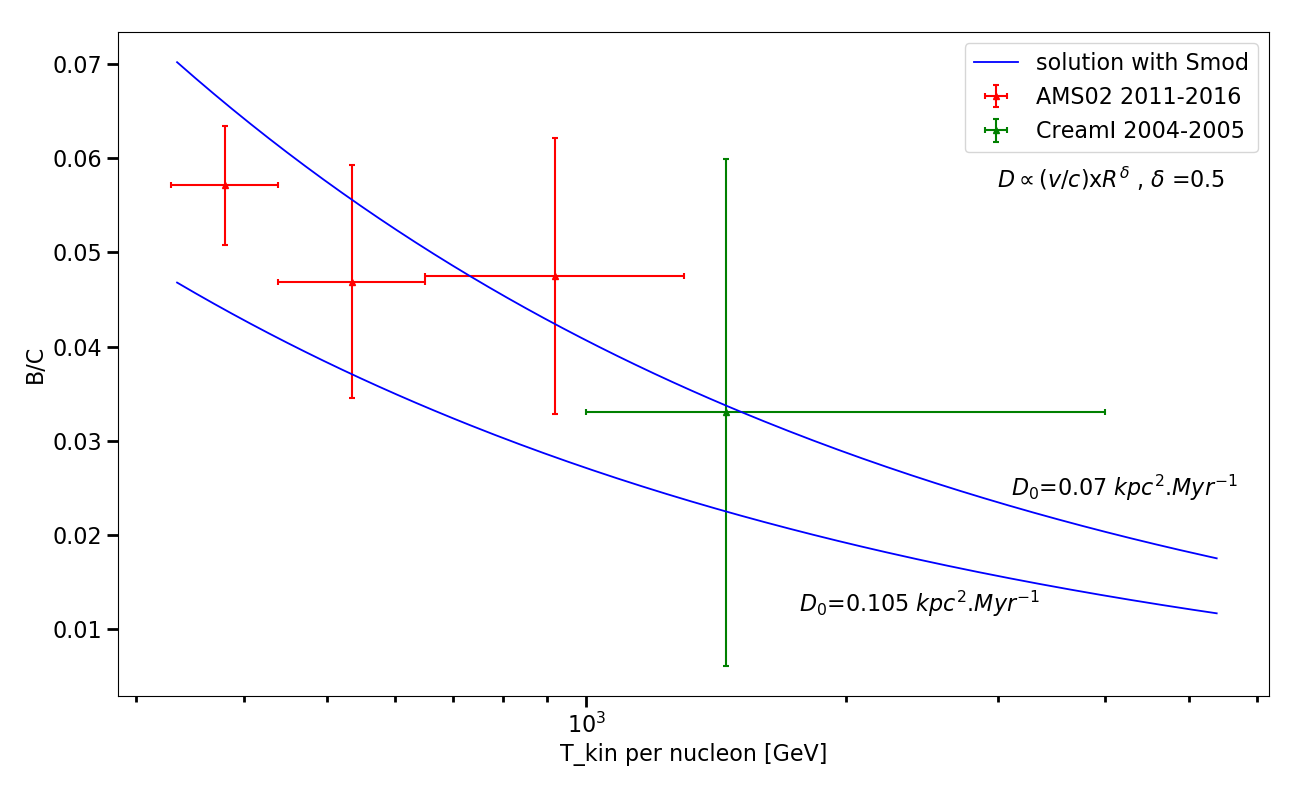}
\caption{Boron to Carbon (B/C) ratio, at high energy ($K\geq 200$ GeV), at Earth, vs. the kinetic energy per nucleon $K/n$, for $L=10$ kpc, $\delta = 0.5$ (Kraichnan turbulence) and two values of the parameter $D_0$.}
\label{fig:D0}
\end{figure}

In addition to reproducing the flux of primary GCR nuclei, it is instructive to investigate whether the observed ratios of secondary nuclei to primary nuclei (so called secondary-to-primary ratios) and secondary radioactive nuclei to secondary stable nuclei ratios could also be matched by  a central source of cosmic rays. This is addressed in the next sections.

\subsection{Secondary-to-primary ratios} 
\label{subsec:BOnC}

From the secondary-to-primary ratio ($\mathrm{II}/\mathrm{I}$), one classically constrains the diffusion coefficients parameters (\cite{genolini2015}) using the high energy part of the ratio. Indeed, in that energy range where energy losses can be neglected, the ratio ($\mathrm{II}/\mathrm{I}$) scales as $D_\mathrm{I}/D_\mathrm{II}^2$ where $D_\mathrm{I}$ and $D_\mathrm{II}$ are respectively the diffusion coefficient of species $\mathrm{I}$ and $\mathrm{II}$ at the same given kinetic energy per nucleon $K/n$ (see Sect.~\ref{subsec::secondary}).

In our case, the energy domain in which energy losses are inefficient and the diffusion coefficient can be estimated directly corresponds to $K/n \gtrsim 10^2$ GeV, where the observational data become scarse and have larger uncertainties. We can nevertheless estimate the normalisation coefficient, $D_0$, to be on the order of 0.07–0.1~kpc$^2$/Myr, should the GCCR Boron-to-Carbon (B/C) ratio at Earth be matching the observed one at high energy (see Fig.~\ref{fig:D0}). 

\begin{figure*}[ht!]
\centering
\includegraphics[width=0.48\textwidth]{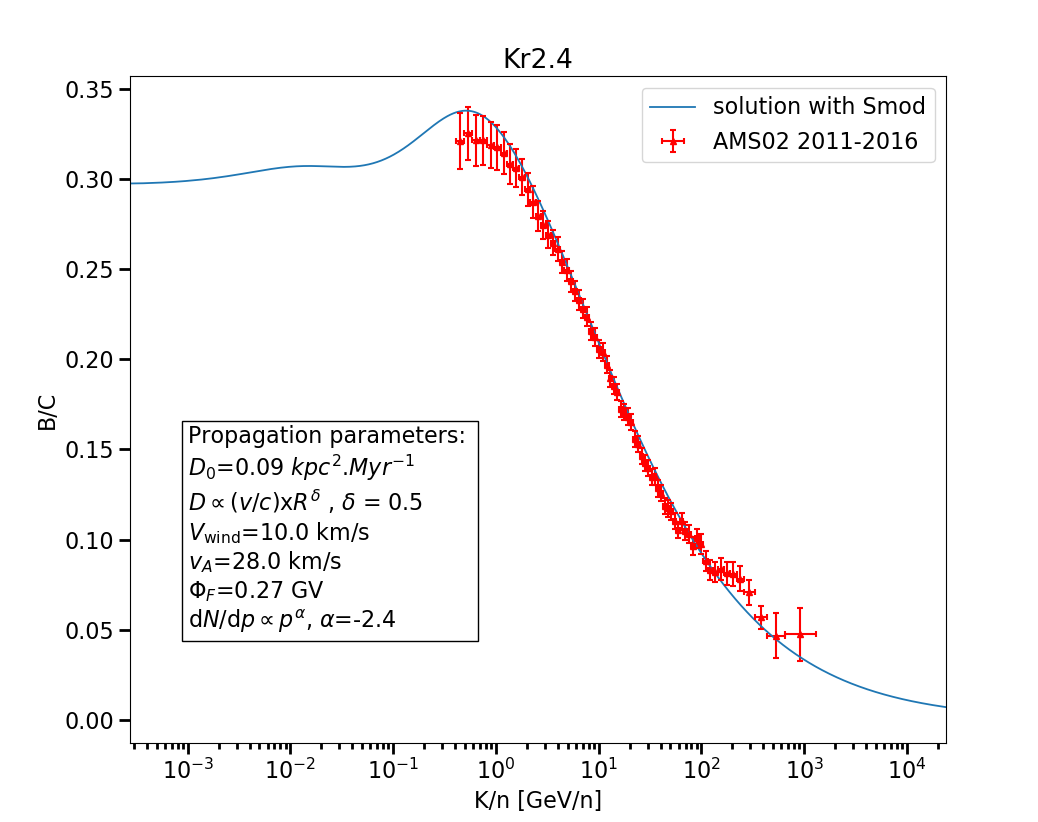}
\includegraphics[width=0.48\textwidth]{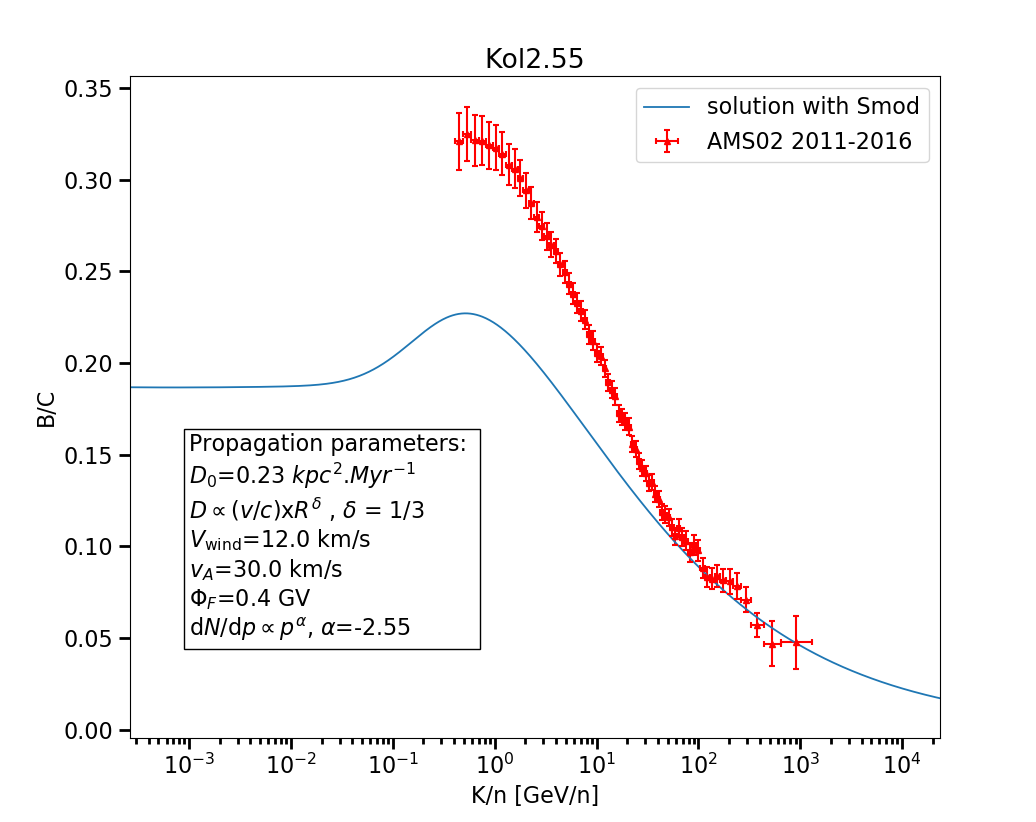}
\caption{Boron to Carbon (B/C) ratio at Earth against kinetic energy per nucleon $K/n$ for the Kr2.4 (left) and Kol2.55 (right) models (see table \ref{table:power}).}
\label{fig:BsurC}
\end{figure*}

Once $D_0$ is fixed in this way, the B/C ratio of the GCCRs at Earth can be obtained over the entire energy range, under the assumption that the other parameters are chosen, say, to reproduce the observed primary spectra (e.g. protons and carbon nuclei). The results are displayed on Fig.~\ref{fig:BsurC}, where we show the B/C ratio as a fuction of kinetic energy per nucleon, for the models Kr2.4 and Kol2.55, corresponding to a Kraichnan-like and Kolmogorov-like energy scaling of $D(K)$, respectively (see Table~\ref{table:power}). As can be seen, if one adopts the source spectral indices, resp. 2.4 and 2.55, which lead to primary spectra matching the observed ones, the resulting evolution of the B/C ratio with energy is very different from the observed one in the case of a Kolmogorov-like scaling (Kol2.55 model), but remarkably close to it in the case of a Kraichnan-like scaling (Kr2.4 model).

\subsection{\textbf{"Cosmic-ray clocks"}}

Another important ingredient of GCR phenomenology is related to the so-called cosmic-ray clocks, i.e. radioactive secondary nuclei with half-lives comparable with the GCR dynamical timescales, which can therefore provide some information about the timescale(s) of cosmic-ray transport in the Galaxy (typical age, confinement or homogenization time, depending on the model). The key observable, in the respect, is the radioactive-to-stable isotopic ratio for appropriate secondary nuclei. A classical example is the $^{10}$Be/$^9$Be ratio: both isotopes are produced exclusively by (inverse) spallation during CR transport, and $^{10}$Be has a half-life of $\tau_{1/2} \simeq 1.4$~Myr, while $^{9}$Be is stable.

Qualitatively, it is easy to see how such cosmic-ray clocks can be used in standard GCR phenomenological models to constrain, for instance, the size of the magnetic halo. At low energy, the GCR confinement time is large enough (and the relativistic time dilation effect weak enough) that most $^{10}$Be nuclei produced in flight decay before escaping the Galaxy. The resulting $^{10}$Be/$^9$Be ratio is thus expected to be "low", at a level that depends on the size of the halo, $L$: for thicker halos, more radioactive nuclei will be able to decay before escaping, thereby reducing the radioactive-to-stable ratio (see e.g. \cite{Strong1998}). At higher energy, the confinement time is reduced and the time dilation effect increases, so that for sufficiently large energies, the observed ratio is essentially expected to match the production ratio, taken to be the same as in \cite{Strong1998} for example. The lack of precise measurements over a sufficient large energy range, however, prevents an accurate determination of the halo size. Thus, the $L/D$ degeneracy cannot totally broken, as illustrated by benchmark models that are still in use. For a more detailed description, see the review of \cite{strong2007cosmic} and references therein.

The above behaviour of radioactive-to-stable ratios is of course expected to be similarly observed in the case of GCCRs. It would be misleading to expect that the resulting $^{10}$Be/$^9$Be ratio be significantly lower in the case of GCCRs, because the typical propagation time from the central source to the Earth vicinity is longer than the typical age of cosmic-rays reaching us from more nearby sources in the case of distributed GCR source scenarios (spanning the Galactic disk). However, both Be isotopes are secondary, and even in the case of GCCRs, most of those who are actually observed on Earth are still produced by spallation in our local part of the Galactic disk, at a rate and level that essentially depends on the nearby ISM density and local GCR flux, i.e. not very different from what occurs in the case of distributed sources.

Quantitatively, it is easy to extend our approach, as described in sections \ref{sec:model} and \ref{sec:ResolutionScheme}, to compute the distribution of secondary radioactive nuclei as well: one simply needs to add a term $-\Gamma_\mathrm{rad} \Psi$ in the right hand side of Eq.~(\ref{eq:DiffusionEquation}), where $\Gamma_\mathrm{rad} = \mathrm{ln}(2)/(\gamma \, \tau_{1/2})$ is the decay constant, taking into account time dilation for radioactive nuclei with Lorentz factor $\gamma = \gamma(K)$.

\begin{figure*}[ht!]
\centering
\includegraphics[width=0.48\textwidth]{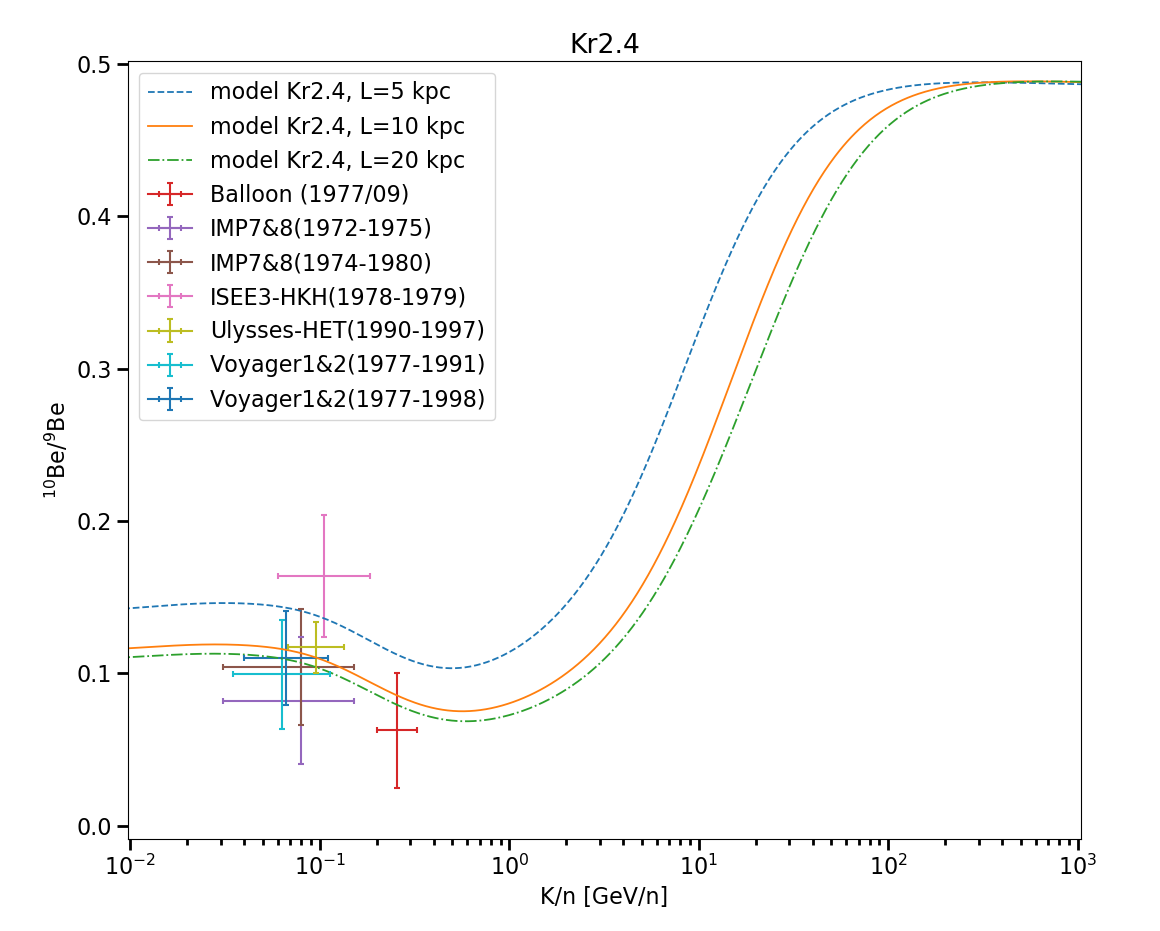}
\includegraphics[width=0.48\textwidth]{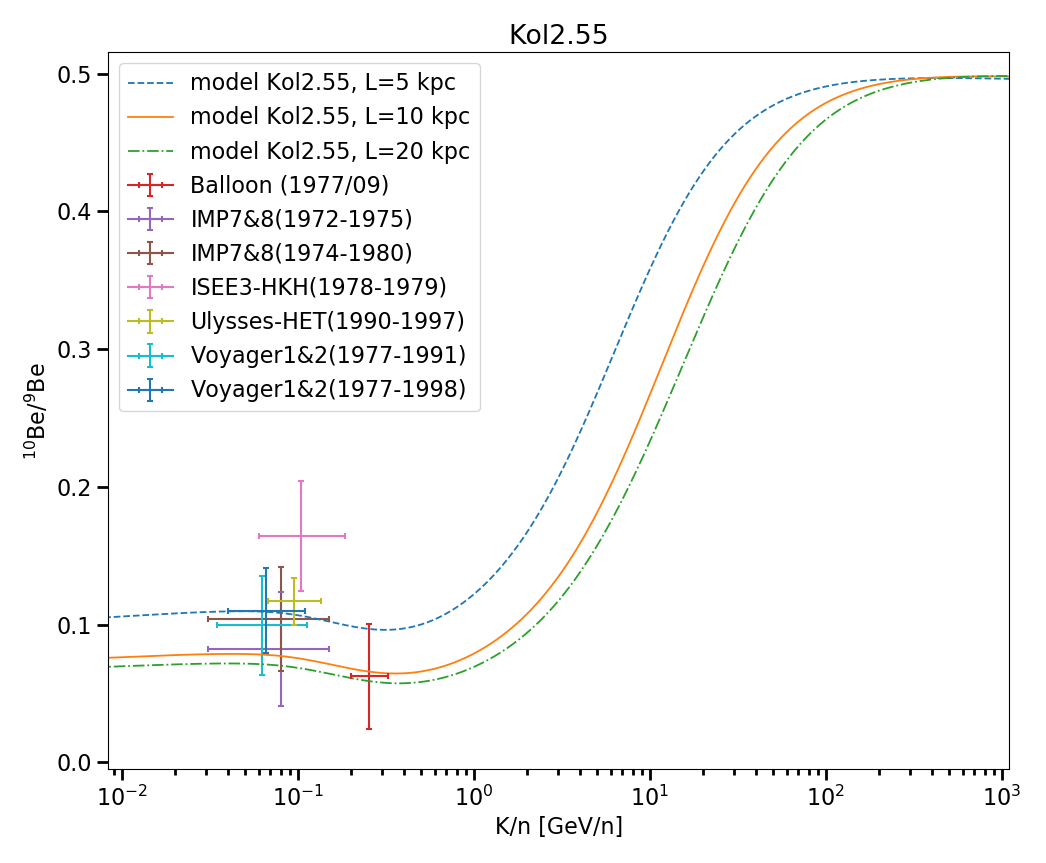}
\caption{$^{10}$Be/$^9$Be ratio at Earth as a function of kinetic energy per nucleon $K/n$ for the Kr2.4 (left) and Kol2.55 (right) models. The different curves correspond to different halo size, $L = 5$, 10 and 20~kpc, as indicated (the other parameters are fixed as in Table~\ref{table:power}).}
\label{fig:BeonBe}
\end{figure*}

The results are shown in Fig.~\ref{fig:BeonBe} for the two models Kr2.4 and Kol2.55 models and halo sizes $L = 5$, 10 and 20~kpc. As can be seen, $^{10}$Be/$^9$Be ratios similar to the observed ones are obtained, for halo sizes $L \gtrsim 4$ kpc, in conformity with the results of Sect.~\ref{subsubsec:halosize}. This confirms that a significant fraction of GCCRs could be present among the local cosmic-rays without disturbing significantly the familiar phenomenology.

\section{Summary and discussion}
\label{sec:discussion}

In the absence of a definite model that would be supported by clear evidence, it appears important to revisit in a broader perspective what is often considered common knowledge regarding cosmic ray origin and propagation in the Galaxy. Most of the GCR phenomenological studies assume that cosmic ray sources are distributed throughout the Galatic disk, with a space and time granularity that can be neglected, at first order, with respect to the typical source seperation distances and repetition times. While this is a reasonable assumption, especially since such distributed sources are indeed known (notably SNRs and superbublles), it does not exclude that other types of sources may also contribute to the GCRs.

In the framework of GCR studies assuming distributed sources, reasonable estimates of the GCR diffusion coefficient and confinement time could be obtained, which in turn provided an estimate of the required GCR source power, comparable to a fair fraction of the total power released by SN explosions in the Galaxy. It turns out, however, that similar or even possibly larger average power is now known to be released around the Galactic center through episodic events, and that these events lead to astrophysical conditions that appear to be potentially propitious for particle acceleration. According to \cite{guo2012fermi} and \cite{miller2016interaction}, the event that led to the so-called \textit{Fermi bubbles} released a total energy of $10^{57}$~ergs, for an average injection rate of 1--7 $10^{42}$ erg/s, which exceeds the total kinetic power of SN explosions in the interstellar medium by large amounts. It is thus natural to investigate whether such events could play a role in the global GCR phenomenon, and if so how the phenomenology would be affected.

The main result of this paper is the confirmation that, with the same type of parameters as those derived from distributed-sources GCR studies, the contribution of the GCCRs cannot be a priori neglected. We found, indeed, that save for models with very thin halos (e.g. MIN), the required injection power to match the observed GCR fluxes ($\mathcal{P}_\mathrm{inj} \lesssim 10^{41}$ erg/s) is only a fraction of the inferred available power. For some transport parameters, the GCCRs could even dominate the GCR flux over a large range of energies. If this is the case, then they will have to satisfy the observational constraints directly. To see if this could be possible, we investigated the caracteristics of the GCCRs as if they were alone in the Galaxy, and found that, even with simple generic assumptions and first order modelling (see Sect.~\ref{sec:model}), it is indeed possible to reproduce the locally observed primary CR spectra as well as secondary-to-primary ratios (B/C) and secondary radioactive-to-stable ratio ($^{10}$Be/$^9$Be), although the latter is only true for a Kraichnan-like energy dependence of the diffusion coefficient.

However, even a subdominant (but non negligible) contribution of the GCCRs may have important consequences on the general CR phenomenology. In particular, the source composition and source spectrum of the GCCRs may be expected to be somewhat different from those of the other sources. In regions of the spectrum where both contributions are roughly similar in magnitude, this may lead to gradual changes in the composition (hence different spectra for different nuclei), or to specific features in the elemental spectra. Such effects will be studied in a forthcoming paper.

If the GCCRs are able to contribute at a non negligible level, some of the conclusions of the standard approach regarding the transport parameters of the energetic particles in the Galaxy may also have to be revised, which can result in the relaxation of some of the usual constraints. In particular, the modelling of the multi-wavelength emission of SNRs may not have to be done with the requirement that the associated energetic particles be also compliant with the entire GCR phenomenology, including the maximum energy problem or the elemental and isotopic ratios. Likewise, it may be possible to relax the constraints associated with the study of the so-called cosmic-ray clocks (radioactive secondary nuclei), if two main components with different timescales are mixed together among the GCRs.

The fact that GCCR typically need more time to reach the Solar System than the average cosmic ray from more evenly distributed sources has also some consequences on the link between CR nuclei and leptons. Since electrons and positrons rapidly lose energy as they propagate in the Galactic  magnetic field, those observed at Earth must have been generated relatively nearby. GCCRs can only contribute secondary $e^{\pm}$, e.g. produced by collisions of high energy protons and helium nuclei on the atoms of the ISM (see e.g. \cite{delahaye2009galactic}), while the bulk of the energetic leptons should still be due to local Galactic sources. Of course, such a ``decoupling" between nuclei and lepton sources is not specific to our approach, and is also suggested for instance in the context of the observed positron anomaly (\cite{adriani2009anomalous}) where nearby pulsar sources can be invoked (\cite{hooper2009pulsars}, \cite{profumo2012dissecting}, \cite{linden2013probing}). Note that this decoupling may also alleviate some of the problems that arise when trying to reconcile observations with models of dark matter (\cite{boudaud2015new}, \cite{cirelli2009model}).

In this paper, we have estimated the average contribution of the GCCRs in the vicinity of the solar system, assuming steady state. Because of the very nature of the potential GCCR sources, the assumption of a constant particle injection is clearly wrong. However, for not too high energies, i.e. as long as the ``confinement time" of the particles is much larger than the repetition time between acceleration events, the steady state solution provides an acceptable approximation. 

To estimate the energy range where the steady state assumption can be expected to be valid, we can compare the relevant timescales. The most recent major event in the Galatic center is suggested to have occurred $\sim 3$ Myr ago (\cite{miller2016interaction}) and to have lasted for $\sim 0.1 - 0.5$ Myr (\cite{guo2012fermi}), leading to the so-called Fermi Bubbles (\cite{acero2016development}). We may thus assume a repetition time of the order of a few Myr. For each event, GCCR injection occurs on a much shorter timescale and can be approximated here as being instantaneous. This gives rise to a diffusion front propagating outwards from the Galactic center. If diffusion were isotropic in a homogeneous medium with diffusion coefficient $D$, the density of particles at distance $r$ from the source, at a time $t$ after injection, would be simply given by the Green function:
\begin{equation}
\Psi_G(\vec{r},t>0) = \frac{N_0}{(4\pi D t)^{3/2}} \exp\left\lbrace-\frac{\vec{r}\,^2}{4Dt}\right\rbrace,
\label{eq:PsiGreenInf}
\end{equation}
where $N_0$ is the number of particles injected at $\vec{r} = 0$ at $t=0$.

At any given position $r$, this density rapidly increases to a maximum and decreases more slowly as the diffusion sphere expands. The typical duration of the event, as seen at radius $r$, can be estimated as the time during which the particle density is larger than half of its maximum value: $\Delta t_{1/2} \simeq 0.45 r^2/D$.

Thus, at radius $r$, the steady-state solution provide a good approximation of the actual GCCR flux at any given time up to an energy $K$ such that $D(K) < 0.45 r^2/\Delta t_\mathrm{s}$, where $\Delta t_\mathrm{s}$ is the typical time interval between two source episodes. Numerically, at the solar radius, this gives:
\begin{equation}
D(K) \la (12 \,\mathrm{kpc}^2/\mathrm{Myr}) \times \left(\frac{\Delta t_\mathrm{s}}{3\,\mathrm{Myr}}\right)^{-1} \times \left(\frac{r}{r_\odot}\right)^2.
\label{eq:Dmax_steady}
\end{equation}

With the parameterization of the diffusion coefficient adopted above, $D = D_{0} \, \beta \, R_{\mathrm{GV}}^{\delta}$, with $D_0 \sim 0.07$--0.1~kpc$^2$/Myr, this corresponds to rigidities
\begin{equation}
R \la (200^{1/\delta} GV) \times \left(\frac{\Delta t_\mathrm{s}}{3\,\mathrm{Myr}}\right)^{-1/\delta} \times \left(\frac{r}{r_\odot}\right)^{2/\delta}.
\label{eq:Rmax_steady}
\end{equation}
In particular, for the Kr2.4 model, the above solution is roughly valid up to $\sim 40$~TeV, and for the Kol2.55 model, up to 8~PeV.

At higher energy, however, the intermittent nature of the central source will affect the main characteristics of the GCCR. Qualitatively, one should expect a reduction of the high energy particles at any time when the diffusion front from the last event at these energies has already passed the solar radius, producing a knee-like feature. Ankle-like features can also be obtained at energies where the diffusion front from the last event is arriving at the solar radius at the time of observation, while the previous one has already left. These effects will be studied in more detail together with associated composition features in a forthcoming paper \citep{JaupartTransient}. 

The previous considerations can be extended to any position inside the Galactic disk. For any galactocentric radius $r<20$ kpc, there is a critical rigidity $R_\mathrm{crit}$ (given by Eq. \ref{eq:Rmax_steady}) above which the steady state approximation is no longer valid. More specifically, in the inner region of the Galaxy, at $r \simeq 1$ kpc for example, this yields $R_\mathrm{crit}  \sim 2.8^{1/\delta} $ GV,  corresponding to a few GeV of kinetic energy in the Kr2.4 and Kol2.55 models. The diffusive front from the last event has thus gone through the inner regions of the Galaxy ($r \lesssim 1$~kpc) for GCCRs with a kinetic energy above the GeV level. Thus, these regions are depleted in these GCCRs, which acts to flatten the distribution given by Fig. \ref{fig:halo}. This is also the energy range which dominates the production of gamma-rays from $\pi^0$ decay. Therefore, in order to compute the gamma-ray background associated with the GCCRs and compare it to the observation, a more detailed, time-dependent treatment is needed. This will be addressed in a separate paper.

\begin{acknowledgements}
      The authors are grateful to Pierre Salati for his useful advice and comments. They also acknowledge support from Agence Nationale de la Recherche (grant ANR-17-CE31-0014). This work is partially supported by an ENS scholarship to EJ from the University of Lyon.
\end{acknowledgements}

\begin{appendix}

\section{Numerical procedure to account for energy losses}
\label{AP::1}

We start from Eq.~(\ref{eq:Pi0}) and define $\tau_{D,j}$ and $\tilde{Q}_j$ by
\begin{eqnarray}
\tau_{D,j}^{-1}= \Gamma_{ISM}+\frac{1}{2h} \left\lbrace V_\mathrm{wind} + D\,S_j\,\coth\left( S_j \, L/2 \right) \right\rbrace \\
\tilde{Q}_j(K) = \frac{\mathrm{d}N/\mathrm{d}K}{2\pi h R^2 J_1(u_j)^2}.
\end{eqnarray}
We then re-write Eq.~(\ref{eq:Pi0}) to obtain:
\begin{equation}
\alpha_j \overline{P}_j = \alpha_j P_j + \frac{d}{dx} \left\lbrace b^{loss} P_j - \gamma \frac{d}{dx}P_j \right\rbrace,
\end{equation}
where $\alpha_j=K/\tau_{D,j}$, $\overline{P}_j= \tau_{D,j} \times \tilde{Q}_j$, $x=\log(K)$ and $\gamma = \chi(K)/K$.

We then discretize the previous equation on a logarithmic scale with $N+1$ values between $K_{\min}$ and $K_{\max}$ with step of $ \Delta x = \frac{1}{N} \log \left\lbrace \frac{K_{\max}}{K_{\min}} \right\rbrace$. Hence the $k$-th value on that grid is $x_k=\log\left(K_{\min} \right) + k\, \Delta x$. We obtain the discretized equation:
\begin{eqnarray}
\alpha_{j,k} \overline{P}_{j,k} = \left\lbrace - \frac{b^{loss}_{k-1}}{2\Delta x} - \frac{\gamma_{k-1/2}}{\Delta x^2} \right\rbrace P_{j,k-1} 
 \nonumber \\
 + \left\lbrace \alpha_k + \frac{\gamma_{k+1/2}+\gamma_{k-1/2}}{\Delta x^2} \right\rbrace P_{j,k} + \left\lbrace \frac{b^{loss}_{k+1}}{2\Delta x} - \frac{\gamma_{k+1/2}}{\Delta x^2} \right\rbrace P_{j,k+1},
\end{eqnarray}
that is:
\begin{eqnarray}
\overline{P}_{j,k} = a_k P_{j,k-1} + b_k P_{j,k} + c_k P_{j,k+1} \\
a_k = \frac{1}{\alpha_k} \left\lbrace - \frac{b^{loss}_{k-1}}{2\Delta x} - \frac{\gamma_{k-1/2}}{\Delta x^2} \right\rbrace \\
b_k=\frac{1}{\alpha_k} \left\lbrace \alpha_k + \frac{\gamma_{k+1/2}+\gamma_{k-1/2}}{\Delta x^2} \right\rbrace \\
c_k = \frac{1}{\alpha_k} \left\lbrace \frac{b^{loss}_{k+1}}{2\Delta x} - \frac{\gamma_{k+1/2}}{\Delta x^2} \right\rbrace,
\end{eqnarray}
namely:
\begin{equation}
[\overline{P}_j]=M(a,b,c) [P_j]
\end{equation}
where $M(a,b,c)$ is a matrix to be inverted. 

Regarding the boundary conditions, we need to specify them on our energy grid to perform the inversion. At high kinetic energy, we expect that energy losses and diffusive reacceleration are not significant so that: 
\begin{equation}
P_{j,N} = \overline{P}_{j,N}
\end{equation}
At low energy, we note that the primary cosmic-ray data can be fitted by power law spectra, and thus impose this as an \emph{ad hoc} boundary condition, which can be expressed as:
\begin{equation}
P_{j,1}-P_{j,0} = P_{j,0} - P_{j,-1}
\end{equation}
This allows to invert the relation $[\overline{P}_j]=M(a,b,c) [P_j]$ to obtain each Bessel order $P_j$.

\end{appendix}

\bibliographystyle{aa}
\bibliography{article_apc}

\end{document}